# Theoretical Error Performance Analysis for Variational Quantum Circuit Based Functional Regression

**Jun Qi**
Georgia Institute of Technology
jqi41@gatech.edu

**Chao-Han Huck Yang**
Georgia Institute of Technology
huckiyang@gatech.edu

**Pin-Yu Chen**
IBM Research
pin-yu.chen@ibm.com

**Min-Hsiu Hsiuh**
Hon Hai (Foxconn) Research Institute
minhsiuh@gmail.com

## Abstract

The noisy intermediate-scale quantum (NISQ) devices enable the implementation of the variational quantum circuit (VQC) for quantum neural networks (QNN). Although the VQC-based QNN has succeeded in many machine learning tasks, the representation and generalization powers of VQC still require further investigation, particularly when the dimensionality of classical inputs is concerned. In this work, we first put forth an end-to-end quantum neural network, TTN-VQC, which consists of a quantum tensor network based on a tensor-train network (TTN) for dimensionality reduction and a VQC for functional regression. Then, we aim at the error performance analysis for the TTN-VQC in terms of representation and generalization powers. We also characterize the optimization properties of TTN-VQC by leveraging the Polyak-Lojasiewicz (PL) condition. Moreover, we conduct the experiments of functional regression on a handwritten digit classification dataset to justify our theoretical analysis.

## 1 INTRODUCTION

The imminent of quantum computing devices opens up new possibilities for exploiting quantum machine learning (QML) [1, 2, 3] to improve the efficiency of classical machine learning algorithms in many new scientific domains like drug discovery [4] and efficient solar conversion [5]. Although the exploitation of quantum computing devices to carry out QML is still in its early exploratory states, the rapid development in quantum hardware has motivated advances in quantum neural network (QNN) to run in noisy intermediate-scale quantum (NISQ) devices [6, 7, 8, 9, 10], where not enough qubits could be spared for quantum error correction and the imperfect qubits have to be directly employed at the physical layer [11, 12, 13]. Even though, a compromised QNN is proposed by employing a quantum-classical hybrid model that relies on an optimization of the variational quantum circuit (VQC) [14, 15]. The resilience of the VQC to certain types of quantum noise errors and the high flexibility concerning coherence time and gate requirements admit VQC to apply to many promising applications on NISQ devices [16, 17, 18, 19, 20, 21, 22, 23].

Although many empirical studies of VQC for quantum machine learning have been reported, its theoretical understanding requires further investigation in terms of representation and generalization powers, particularly when the non-linear operator is employed for dimensionality reduction. This work introduces a tensor-train network (TTN) on top of the VQC model to implement a TTN-VQC. The TTN is a non-linear operator mapping high-dimensional features into low-dimensional ones. Then, the resulting low-dimensional features go through the framework of VQC. Compared with a hybrid model where the operation of dimensionality reduction is constituted by a classical neural network (NN) [24], TTN can be genuinely realized by utilizing universal quantum circuits [25, 26, 19], and an end-to-end quantum neural network can be truly set up.

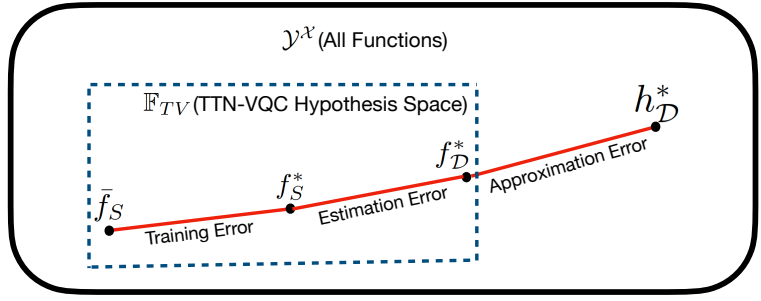

Figure 1: *An illustration of error decomposition technique.* $h_{\mathcal{D}}^*$ *is a smooth target function in a family of all functions* $\mathcal{Y}^{\mathcal{X}}$ *over a data distribution* $\mathcal{D}$; $\mathbb{F}_{TV}$ *denotes the family of TTN-VQC operators as shown in the dashed square;* $f_{\mathcal{D}}^*$ *represents the optimal hypothesis from the space of TTN-VQC operators over the distribution* $\mathcal{D}$; $f_S^*$ *denotes the best empirical hypothesis over the set of training samples* $S$; $\bar{f}_S$ *is the returned hypothesis based on the training dataset* $S$.

In this work, we discuss the theoretical performance of TTN-VQC in the context of functional regression. Functional regression refers to building a vector-to-vector operator such that the regression output can approximate a target operator. In more detail, given a $Q$-dimensional input vector space $\mathbb{R}^Q$ and a measurable $U$-dimensional output vector space $\mathbb{R}^U$, the TTN-VQC-based vector-to-vector regression aims to find a TTN-VQC operator $f : \mathbb{R}^Q \rightarrow \mathbb{R}^U$ such that the output vectors of $f$ can approximate a desirable target one.

In particular, this work concentrates on the error performance analysis for TTN-VQC-based functional regression by leveraging the error decomposition technique [27] to factorize an expected loss over the TTN-VQC operator into the sum of the approximation error, estimation error, and training error. We separately upper bound each error component by harnessing statistical machine learning theory. More specifically, we define $\mathbb{F}_{TV}$ as the TTN-VQC hypothesis space which represents a collection of TTN-VQC operators. Then, given a data distribution $\mathcal{D}$, assuming a smooth target function $h_{\mathcal{D}}^*$ and a set of $N$ training data drawn independent and identically distributed from a data distribution $\mathcal{D}$, for a loss function $\ell$ and an optimal TTN-VQC operator $f_{\mathcal{D}}^* \in \mathbb{F}_{TV}$, an expected loss is defined as:

$$\mathcal{L}_{\mathcal{D}}(f_{\mathcal{D}}^*) := \mathbb{E}_{\mathbf{x} \sim \mathcal{D}} \left[ \ell(h_{\mathcal{D}}^*(\mathbf{x}), f_{\mathcal{D}}^*(\mathbf{x})) \right], \tag{1}$$

which can be minimized by using an empirical loss as:

$$\mathcal{L}_S(f_{\mathcal{D}}^*) := \frac{1}{N} \sum_{n=1}^{N} \ell(h_{\mathcal{D}}^*(\mathbf{x}_n), f_{\mathcal{D}}^*(\mathbf{x}_n)). \tag{2}$$

Since the mean absolute error (MAE) [28] is a 1-Lipschitz continuous [29], the loss function $\ell$ is set as the MAE. Furthermore, we separately define $f_{\mathcal{D}}^*$, $f_S^*$ and $\bar{f}_S$ as an optimal TTN-VQC operator, an empirical optimal operator, and an actual returned operator. Then, as shown in Figure 1, the error decomposition technique [27] factorizes the expected loss $\mathcal{L}_{\mathcal{D}}(\bar{f}_S)$ into three error components as:

$$
\begin{aligned}
\mathcal{L}_{\mathcal{D}}(\bar{f}_S) = & \underbrace{\mathcal{L}_{\mathcal{D}}(f_{\mathcal{D}}^*)}_{blue Approximation\, Error} + \underbrace{\mathcal{L}_{\mathcal{D}}(f_S^*) - \mathcal{L}_{\mathcal{D}}(f_{\mathcal{D}}^*)}_{blue Estimation\, Error} + \underbrace{\mathcal{L}_{\mathcal{D}}(\bar{f}_S) - \mathcal{L}_{\mathcal{D}}(f_S^*)}_{blue Training\, Error} \\
& \leq \mathcal{L}_{\mathcal{D}}(f_{\mathcal{D}}^*) + 2 \sup_{f \in \mathbb{F}_{TV}} |\mathcal{L}_{\mathcal{D}}(f) - \mathcal{L}_S(f)| + \mathcal{L}_{\mathcal{D}}(\bar{f}_S) - \mathcal{L}_{\mathcal{D}}(f_S^*) \\
& \leq \mathcal{L}_{\mathcal{D}}(f_{\mathcal{D}}^*) + 2\hat{\mathcal{R}}_S(\mathbb{F}_{TV}) + \nu,
\end{aligned}
\tag{3}
$$

where $\mathcal{L}_{\mathcal{D}}(f_{\mathcal{D}}^*)$ is associated with the approximation error, $\hat{\mathcal{R}}_S(\mathbb{F}_{TV})$ is an empirical Rademacher complexity [30] over the family $\mathbb{F}_{TV}$, and $\nu$ refers to the training error that results from the optimization bias of gradient-based algorithms. The Rademacher complexity $\hat{\mathcal{R}}(\mathbb{F}_{TV})$ can measure the model complexity and is particularly used for the regression problem [27]. In this work, our theoretical results concentrate on the error analysis by upper-bounding each error component, and our empirical results are illustrated to corroborate our theoretical analysis.

## 1.1 Main Results

Our derived theoretical results in this work and the significance of TTN-VQC-based functional regression are summarized as follows:

- Representation power: our upper bound on the approximation error is derived as $\frac{\Theta(1)}{\sqrt{U}} + \mathcal{O}(\frac{1}{\sqrt{M}})$, where $U$ and $M$ separately denote the number of qubits and the count of quantum measurement. The result suggests that the expressive capability of TTN-VQC can be mainly determined by the number of qubits, and the quality of the expressiveness is also affected by the count of quantum measurements. Larger $U$ and $M$ correspond to the fact that more algorithmic qubits and a longer decoherence time are necessarily required to ensure stronger representation power of TTN-VQC. Furthermore, since more qubits are more likely to result in the problem of Barren Plateaus of VQC during the training process, the introduction of PL condition is significant to handle the problem.

- Generalization power: we derive an upper bound on the estimation error concerning the empirical Rademacher complexity $\hat{\mathcal{R}}_S(\mathbb{F}_{TV})$, which is further upper bounded by a constant as $\frac{2P}{\sqrt{N}}\left(\sqrt{\sum_{k=1}^{K} \Lambda_k^2} + \Lambda'\right)$. Here, $P$, $N$, and $K$ separately denote the input power, the amount of training data, and the order of multi-dimensional tensor; $\Lambda_k$ and $\Lambda$ refer to the upper bounds on the Frobenius norm of TTN parameters. The result of the generalization power suggests that given the training data and model structure, the additive noise corresponds to a larger value of $P$ which results in an upper bound on a weaker generalization capability.

- Optimization bias: the PL condition is employed to initialize the TTN-VQC parameters and the training error can be exponentially converged to a small loss value. The problem of barren plateau is a serious issue in the training process of the quantum neural network [31], especially for a randomized QNN architecture, the variance of gradients exponentially vanishes with the increase of qubits. In this work, we claim that the model setting based on the PL condition could be beneficial to the improvement of the TTN-VQC training.

Besides, our empirical results of functional regression are designed to corroborate the corresponding theoretical results of representation and generalization powers, and the analysis of optimization performance.

## 1.2 Related Work

The related work comprises theoretical and technical aspects. As for the theoretical point, Du *et al.* [32] analyzes the learnability of quantum neural networks with parameterized quantum circuits and gradient-based classical optimizer. A theoretical comparison between this work and Du *et al.* [32] is shown in Table 1, where our theoretical results mainly follow the error decomposition method [27, 33]. More specifically, in this work, we factorize an expected loss based on MAE over a TTN-VQC operator into three error components: approximation error, estimation error, and training error. We separately derive upper bounds on each error component and the results are summarized in Table 1.

| Category | This work | Du *et al.* [32] |
|---|---|---|
| Learning problem | Regression | Classification |
| Dimensionality reduction | TTN | N/A |
| Representation power | $\frac{\Theta(1)}{\sqrt{U}} + \mathcal{O}(\frac{1}{\sqrt{M}})$ | N/A |
| Generalization power | $\frac{2P}{\sqrt{N}}\left(\sqrt{\sum_{k=1}^{K} \Lambda_k^2} + \Lambda'\right)$ | N/A |
| Conditions for Optimization bias | $\mu$-PL + 1-Lipschitz | $\mu$-PL + $\beta$-smooth |

Table 1: A comparison of learning theory for VQC between this work and Du *et al.* [32]

Besides, the techniques of this work rely on the TTN and VQC models. The TTN, also known as matrix product state (MPS) [34], was first put forth by Alexander *et al.* [35] in the applications of machine learning. Chen *et al.* [26] employs MPS to extract low-dimensional features for VQC. Although this work leverages the TTN for dimensionality reduction, we rebuild the TTN as parallel neural network architecture, where the sigmoid activation function is separately imposed upon each neural network. In this work, we choose the TTN for dimensionality reduction for two reasons: (1) Although classical neural networks can be also applied for feature dimensionality [19, 18, 16], the classical-quantum hybrid system may take up more computational resources and it is intractable to place the classical neural networks on quantum devices; (2) numerous works have shown that classical neural networks can be converted into the TTN formats, and the TTN models can maintain or even outperform the classical counterpart [36, 37, 38]. Moreover, since the VQC models have been widely used in the domains of quantum machine learning [39, 40, 41], we follow the standard VQC pipeline such that our theoretical results can be employed for the general VQC model.

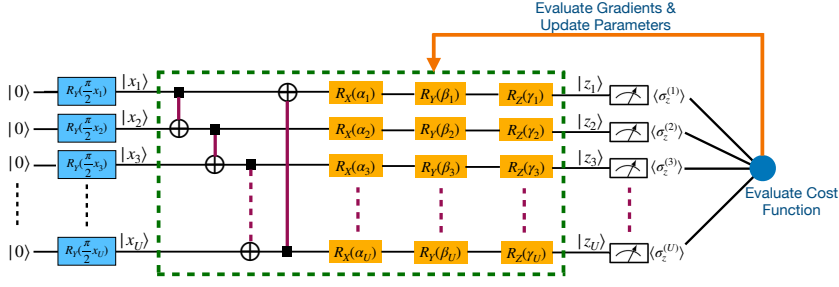

Figure 2: *A VQC model consists of three components: (a) Tensor Product Encoding (TPE); (b) Parametric Quantum Circuit (PQC); (c) Measurement. The TPE employs a series of $R_Y(\frac{\pi}{2}x_i)$ to transform classical data into quantum states. The PQC is composed of CNOT gates and single-qubit rotation gates $R_X$, $R_Y$, $R_Z$ with free model parameters $\boldsymbol{\alpha}$, $\boldsymbol{\beta}$, and $\boldsymbol{\gamma}$. The CNOT gates impose the operation of quantum entanglement among qubits, and the gates $R_X$, $R_Y$, and $R_Z$ can be adjustable during the training stage. To build a deeper model, the PQC model in the green dash square is repeatably copied. The measurement converts the quantum states $|\sigma_z^{(1)}\rangle, |\sigma_z^{(2)}\rangle, ..., |\sigma_z^{(U)}\rangle$ into the corresponding expectation values $\langle\sigma_z^{(1)}\rangle, \langle\sigma_z^{(2)}\rangle, ..., \langle\sigma_z^{(U)}\rangle$. The outputs $|\sigma_z^{(1)}\rangle, |\sigma_z^{(2)}\rangle, ..., |\sigma_z^{(U)}\rangle$ are connected to a loss function and the gradient descent algorithms can be used to update VQC parameters.*

## 2 RESULTS

### 2.1 Preliminaries

Before we delve into the detailed architecture of the TTN-VQC, we first introduce the basic components of TTN and VQC, which have been previously proposed and widely used in quantum machine learning.

#### 2.1.1 Variational Quantum Circuit

As shown in Figure 2, we first introduce a VQC which is composed of three components: (1) Tensor Product Encoding (TPE); (2) Parametric Quantum Circuit (PQC); (3) Measurement.

The TPE model, which is shown in Figure 2 (a), was proposed in [42] and it aims at converting a classical data $\mathbf{x}$ into a quantum state $|\mathbf{x}\rangle$ by adopting a one-to-one mapping as:

$$|\mathbf{x}\rangle = \left(\otimes_{i=1}^{U} R_Y(\frac{\pi}{2}x_i)\right)|0\rangle^{\otimes U} = \begin{bmatrix}\cos(\frac{\pi}{2}x_1)\\\sin(\frac{\pi}{2}x_1)\end{bmatrix} \otimes \begin{bmatrix}\cos(\frac{\pi}{2}x_2)\\\sin(\frac{\pi}{2}x_2)\end{bmatrix} \otimes \cdots \otimes \begin{bmatrix}\cos(\frac{\pi}{2}x_U)\\\sin(\frac{\pi}{2}x_U)\end{bmatrix}, \quad (4)$$

where each $x_i$ can be strictly restricted in the domain of $[0, 1]$ such that the conversion between $\mathbf{x}$ and $|\mathbf{x}\rangle$ is a reversely one-to-one mapping.

The PQC framework is illustrated in Figure 2 (b), where $U$ quantum channels are utilized to correspond to currently accessible $U$ qubits on NISQ devices. Here, the controlled-NOT (CNOT) gates realize the quantum entanglement and the single rotation gates $R_X$, $R_Y$, and $R_Z$ compose the PQC model with model free parameters $\boldsymbol{\alpha} = \{\alpha_1, \alpha_2, ..., \alpha_U\}$, $\boldsymbol{\beta} = \{\beta_1, \beta_2, ..., \beta_U\}$ and $\boldsymbol{\gamma} = \{\gamma_1, \gamma_2, ..., \gamma_U\}$. The PQC model corresponds to a linear operator $\mathcal{T}_{\boldsymbol{\theta}_{vqc}}$ that transforms the quantum input state $|\mathbf{x}\rangle$ into the output one $|\mathbf{z}\rangle$. The PQC model in the green dash square is repeatably copied to compose a deeper architecture.

The measurement framework, as shown in Figure 2 (c), outputs expectation values with respect to the Pauli-Z operators, namely $\langle \sigma_z^{(1)} \rangle$, $\langle \sigma_z^{(2)} \rangle$, ..., $\langle \sigma_z^{(U)} \rangle$ which results in the output vector $\mathbf{z} = [\langle \sigma_z^{(1)} \rangle$, $\langle \sigma_z^{(2)} \rangle$, ..., $\langle \sigma_z^{(U)} \rangle]^T$. The expectation vector $\mathbf{z}$ refers to the classical data and it is connected to the operation of functional regression.

### 2.1.2 Tensor-Train Network

TTN refers to a tensor network aligned in a 1-dimensional array and is generated by repetitively singular value decomposition (SVD) [43] to a many-body wave function [25]. To utilize the TTN for dimensionality reduction, in this work, we first define the tensor-train decomposition (TTD) for a 1-dimensional vector and a tensor-train representation for a 2-dim matrix. More specifically, given a vector $\mathbf{x} \in \mathbb{R}^D$ where $D = \prod_{k=1}^{K} D_k$, we reshape the vector $\mathbf{x}$ into a $K$-order tensor $\mathcal{X} \in \mathbb{R}^{D_1 \times D_2 \times \cdots \times D_K}$. Then, given a set of tensor-train ranks (TT-ranks) $\{R_1, R_2, ..., R_{K+1}\}$ ($R_1$ and $R_2$ are set as 1), all elements of $\mathcal{X}$ can be represented by multiplying $K$ matrices $\mathcal{X}_{d_k}^{[k]}$ based on the TT format as:

$$\mathcal{X}_{d_1, d_2, ..., d_K} = \mathcal{X}_{d_1}^{[1]} \mathcal{X}_{d_2}^{[2]} \cdots \mathcal{X}_{d_K}^{[K]} = \prod_{k=1}^{K} \mathcal{X}_{d_k}^{[k]}, \tag{5}$$

where the matrices $\mathcal{X}_{d_k}^{[k]} \in \mathbb{R}^{R_k \times R_{k+1}}$, $\forall d_k \in [D_k]$. The ranks $R_1$ and $R_K$ are set as 1 to ensure the term $\prod_{k=1}^{K} \mathcal{X}_{d_k}^{[k]}$ is a scalar.

Next, we are concerned with the TTD for a 2-dim matrix. A feed-forward neural network with $U$ neurons has the form:

$$\mathbf{y}(u) = \sum_{d=1}^{D} \mathbf{W}(d, u) \cdot \mathbf{x}(d), \quad \forall u \in [U]. \tag{6}$$

If we assume that $U = \prod_{k=1}^{K} u_k$, then we can reshape the 2-order matrix $\mathbf{W}$ as a $D$-order double-indexed tensor $\mathcal{W}$ and it can be factorized into the TT-format as:

$$\mathcal{W}_{(d_1, u_1), (d_2, u_2), ..., (d_K, u_K)} = \mathcal{W}_{d_1, u_1}^{[1]} \mathcal{W}_{d_2, u_2}^{[2]} \cdots \mathcal{W}_{d_K, u_K}^{[K]}, \tag{7}$$

where $\mathcal{W}^{[k]} \in \mathbb{R}^{R_k \times D_k \times U_k \times R_{k+1}}$ is a 4-order core tensor, and each element $\mathcal{W}_{d_k, u_k}^{[k]} \in \mathbb{R}^{R_k \times R_{k+1}}$ is a matrix. Then, we can reshape the input vector $\mathbf{x}$ and the output one $\mathbf{y}$ into two tensors of the same order: $\mathcal{X} \in \mathbb{R}^{D_1 \times D_2 \times \cdots \times D_K}$, $\mathcal{Y} \in \mathbb{R}^{U_1 \times U_2 \times \cdots \times U_K}$, and we build the mapping function between the elements and the input tensor $\mathcal{X}_{d_1, d_2, ..., d_K}$ and the output one $\mathcal{Y}_{u_1, u_2, ..., u_K}$ as:

$$\mathcal{Y}_{u_1, u_2, ..., u_K} = \sum_{d_1=1}^{D_1} \sum_{d_2=1}^{D_2} \cdots \sum_{d_K=1}^{D_K} \mathcal{W}_{(d_1, u_1), (d_2, u_2), ..., (d_K, u_K)} \mathcal{X}_{d_1, d_2, ..., d_K}. \tag{8}$$

Then, by employing TTD for the 1-dim vector $\mathcal{X}_{d_1, d_2, ..., d_K}$ and 2-dim matrix $\mathcal{W}_{(d_1, u_1), (d_2, u_2), ..., (d_K, u_K)}$ separately defined in Eq. (6) and (7), we attain that

$$\begin{aligned}
\mathcal{Y}_{u_1, u_2, ..., u_K} &= \sum_{d_1=1}^{D_1} \sum_{d_2=1}^{D_2} \cdots \sum_{d_K=1}^{D_K} \mathcal{W}_{(d_1, u_1), (d_2, u_2), ..., (d_K, u_K)} \mathcal{X}_{d_1, d_2, ..., d_K} \\
&= \sum_{d_1=1}^{D_1} \sum_{d_2=1}^{D_2} \cdots \sum_{d_K=1}^{D_K} \prod_{k=1}^{K} \mathcal{W}_{d_k, u_k}^{[k]} \odot \prod_{k=1}^{K} \mathcal{X}_{d_k}^{[k]} \\
&= \prod_{k=1}^{K} \sum_{d_k=1}^{D_k} \mathcal{W}_{d_k, u_k}^{[k]} \odot \mathcal{X}_{d_k}^{[k]} \\
&= \prod_{k=1}^{K} \mathcal{Y}_{u_k}^{[k]},
\end{aligned} \tag{9}$$

where $\mathcal{W}_{d_k, u_k}^{[k]} \odot \mathcal{X}_{d_k}^{[k]}$ refers to an element-wise multiplication of the two matrices, and $\sum_{d_k=1}^{D_k} \mathcal{W}_{d_k, u_k}^{[k]} \odot \mathcal{X}_{d_k}^{[k]}$ results in a matrix $\mathcal{Y}_{u_k}^{[k]}$ in $\mathbb{R}^{R_k \times R_{k+1}}$. The ranks $R_1 = R_{K+1} = 1$ ensures the $\prod_{k=1}^{K} \mathcal{Y}_{u_k}^{[k]}$ is a scalar. Based on the framework of TTN, two necessary requirements need to be met as follows: (a) given a $D$-dimensional input vector, we need that $D = \sum_{k=1}^{K} D_k$, where $d_k = [D_k]$ and $R_1 = R_{K+1} = 1$; (b) Given the output vector in $\mathbb{R}^U$, we have $U = \prod_{k=1}^{K} U_k$, where $u_k = [U_k]$, and $R_1 = R_{K+1} = 1$. In particular, the output dimension $U$ in this work corresponds to the number of qubits.

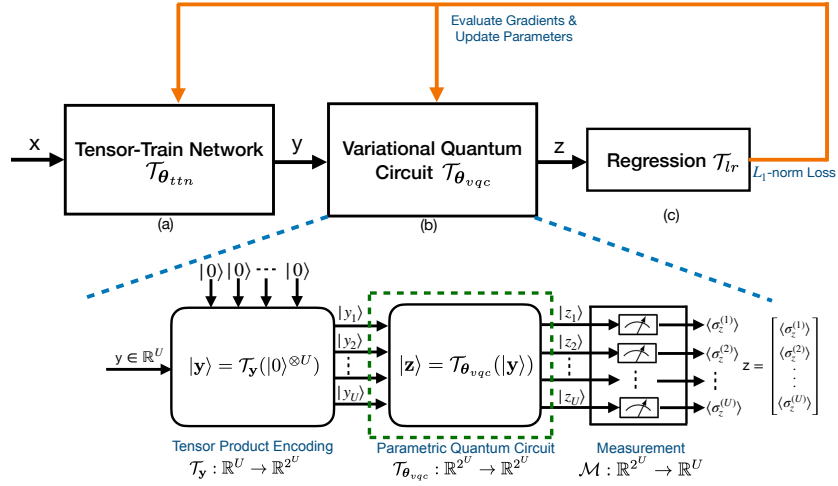

Figure 3: *An illustration of the TTN-VQC architecture. (a) Tensor-Train Network (TTN); (b) Variational Quantum Circuit (VQC); (c) Functional Regression. $\mathcal{T}_{\boldsymbol{\theta}_{ttn}}$ and $\mathcal{T}_{\boldsymbol{\theta}_{vqc}}$ represent the TTN and VQC operators with trainable parameters $\boldsymbol{\theta}_{ttn}$ and $\boldsymbol{\theta}_{vqc}$, respectively. $\mathcal{T}_{\boldsymbol{y}}$ refers to a reversible classical-to-quantum mapping. The VQC model in the green dash square can be repeatably copied to generate a deep parametric model. The framework of functional regression outputs loss values and evaluate gradients of loss functions to update model parameters $\boldsymbol{\theta}_{vqc}$ and $\boldsymbol{\theta}_{ttn}$. $\mathcal{T}_{lr}$ refers to a fixed regression matrix.*

## 2.2 Theoretical Results

This section first exhibits the architecture of TTN-VQC, and then we analyze the upper bounds on the representation and generalization powers and the optimization performance.

### 2.2.1 The Architecture of TTN-VQC

The TTN-VQC pipeline is shown in Figure 3, where (a) denotes the framework of TTN, (b) is associated with the VQC model, and (c) represents the operation of functional regression. The VQC model is based on the standard architecture as shown in 2.1.1, and the TTN is designed according to the framework in 2.1.2. To introduce the non-linearity to the TTN model, a sigmoid activation function $Sigm(\cdot)$ is taken for each $\mathcal{Y}_k(j_k)$ such that

$$\hat{\mathcal{Y}}(j_1, j_2, ..., j_K) = \prod_{k=1}^{K} Sigm(\mathcal{Y}_k(j_k)), \qquad (10)$$

which introduces the non-linearity to the TTN features and corresponds to a parallel neural network structure.

The parallel DNN structure is illustrated in Figure 4, where a $K$-order tensor $\mathcal{X}_{d_1, d_2, ..., d_K}$ is first decomposed into 2-dim matrices $\mathcal{X}_{d_1}^{[1]}$, $\mathcal{X}_{d_2}^{[2]}$, ..., $\mathcal{X}_{d_K}^{[K]}$ and each $\mathcal{X}_{d_k}^{[k]}$ goes through $\mathcal{W}_{d_k, u_k}^{[k]}$. The resulting $\mathcal{Y}_{u_1}^{[1]}$, $\mathcal{Y}_{u_2}^{[2]}$, ..., $\mathcal{Y}_{u_K}^{[K]}$ are non-linearly activated by applying the sigmoid activation function before multiplying them together into a $K$-order tensor $\hat{\mathcal{Y}}_{u_1, u_2, ..., u_K}$. By iterating $u_k \in [U_k]$ and fixing other indices $u_1, u_2, ..., u_{k-1}, u_{k+1}, ..., u_K$, we separately collect a vector associated with the $k^{th}$ order of $\mathcal{Y}$.

More significantly, the non-linearity introduced by the sigmoid function sets up a parallel DNN structure for TTN and helps to build a one-to-one mapping in the TPE framework because the sigmoid function compresses the functional values in the domain of $(0, 1)$. Proposition 1 suggests the sigmoid activation function ensures a one-to-one mapping from the classical data to the quantum state.

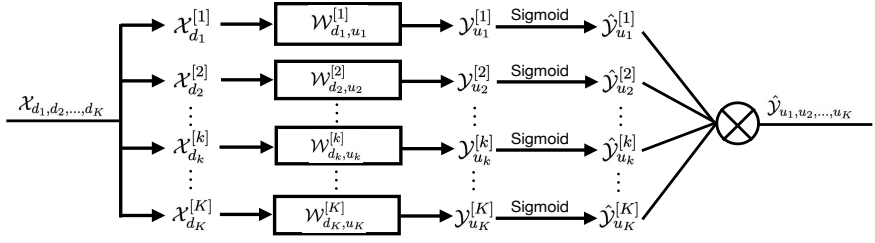

Figure 4: *Reformulating the TTN model in a parallel structure. Each element of the input $K$-order tensor $\mathcal{X}_{d_1, d_2, \ldots, d_K}$ is factorized into $K$ matrices $\mathcal{X}_{d_k}$ by utilizing TTD. Each $\mathcal{X}_{d_k}^{[k]}$ goes through the TTN associated with model parameters $\mathcal{W}^{[k]}$. The sigmoid function is imposed upon the output $\mathcal{Y}_{u_k}^{[k]}$, and all $\mathcal{Y}_{u_k}^{[k]}$ are all multiplied to form the output $\hat{\mathcal{Y}}_{u_1, u_2, \ldots, u_K}$.*

**Proposition 1.** *The sigmoid activation function applied to the TTN ensures the TPE as a linear unitary operator $|\mathbf{y}\rangle = \mathcal{T}_{\mathbf{y}}(|0\rangle^{\otimes U})$ such that a quantum state $|\mathbf{y}\rangle$ can be generated from a classical vector $\mathbf{y}$. On the other hand, the classical vector $\mathbf{y}$ can be exactly deduced based on the operator $\mathcal{T}_{\mathbf{y}}$.*

Proposition 1 can be justified based on Eq. (4), where $\cos(\frac{\pi}{2}x_i)$ and $\sin(\frac{\pi}{2}x_i)$ are reversible one-to-one functions because of each $x_i \in (0, 1)$. Then, we can deduce the original classical vector $\mathbf{y}$ given the quantum state $|\mathbf{y}\rangle$.

The VQC outputs a classical vector $\mathbf{z} = [\langle \sigma_z^{(1)} \rangle, \langle \sigma_z^{(2)} \rangle, \ldots, \langle \sigma_z^{(K)} \rangle]^T$, and then $\mathbf{z}$ is connected to the framework of functional regression, where a fixed linear regression operator $\mathcal{T}_{lr}$ further transforms $\mathbf{z}$ into the output vector. The MAE is taken to measure the loss value and the related gradients of the loss function, which are used to update the parameters of both VQC and TTN models.

### 2.2.2 Upper Bounds On the Approximation Error

Theorem 1 shows an upper bound on the approximation error. The upper bound on the approximation error relies on the theoretical analysis of the inherent parallel structure for the TTN model and the universal approximation theory utilized for neural networks [44, 45, 46]. Theorem 1 suggests that the representation power of linear operator $\mathcal{M} \circ \mathcal{T}_{\boldsymbol{\theta}_{vqc}} \circ \mathcal{T}_{\mathbf{y}}$ is strengthened by applying a non-linear operator $\mathcal{T}_{\boldsymbol{\theta}_{ttn}}(\mathbf{x})$.

**Theorem 1.** *Given a smooth target function $h_{\mathcal{D}}^* : \mathbb{R}^Q \to \mathbb{R}^U$ and a classical data $\mathbf{x}$, there exists a TTN-VQC $g(\mathbf{x}; \boldsymbol{\theta}_{vqc}, \boldsymbol{\theta}_{ttn}) = \mathcal{M} \circ \mathcal{T}_{\boldsymbol{\theta}_{vqc}} \circ \mathcal{T}_{\mathbf{y}} \circ \mathcal{T}_{\boldsymbol{\theta}_{ttn}}(\mathbf{x})$, we obtain*

$$\mathcal{L}_{\mathcal{D}}(f_{\mathcal{D}}^*) = \|h_{\mathcal{D}}^*(\mathbf{x}) - \mathcal{T}_{lr}\left(\mathbb{E}\left[g(\mathbf{x}; \boldsymbol{\theta}_{vqc}, \boldsymbol{\theta}_{ttn})\right]\right)\|_1 \leq \frac{\Theta(1)}{\sqrt{U}} + \mathcal{O}\left(\frac{1}{\sqrt{M}}\right), \tag{11}$$

*where $U$ and $M$ separately refer to the number of qubits and the count of quantum measurement, and $\mathbb{E}[g(\mathbf{x}; \boldsymbol{\theta}_{vqc}, \boldsymbol{\theta}_{ttn})]$ represents an expectation value of the output measurement.*

The upper bound in Eq. (11) implies that the number of qubits $U$ and the count of measurement $M$ jointly decide the representation power of TTN-VQC, and larger values of $U$ and $M$ are expected to lower the upper bound. However, a larger value $U$ requires an advanced quantum computer with more logic qubits, but more qubits are likely to degrade the optimization performance because of the problem of Barren Plateaus. To strike a balance between a large number of qubits and low optimization bias, the PL condition is introduced to initialize the TTN-VQC model.

### 2.2.3 Upper Bounds on the Estimation Error

Theorem 2 suggests the upper bounds on the estimation error. The upper bound on the estimation error can be derived based on the empirical Rademacher complexity $\hat{\mathcal{R}}_S(\mathbb{F}_{TV})$, which is defined as:

$$\hat{\mathcal{R}}_S(\mathbb{F}_{TV}) := \mathbb{E}_{\boldsymbol{\epsilon}}\left[\sup_{f \in \mathbb{F}_{TV}} \frac{1}{N} \sum_{n=1}^N \epsilon_n f(\mathbf{x}_n)\right], \tag{12}$$

where $N$ samples $S = \{\mathbf{x}_1, \mathbf{x}_2, ..., \mathbf{x}_N\}$, and $\boldsymbol{\epsilon} = \{\epsilon_1, \epsilon_2, ..., \epsilon_N\}$ refers to a set of $N$ Rademacher random variables taking on values 1 and $-1$ with an equal likelihood. The empirical Rademacher complexity measure how well the functional family $F_{TV}$ correlates with random noise $\boldsymbol{\epsilon}$ on the dataset $S$, and it describes the richness of the family $F_{TV}$: a richer family $F_{TV}$ can generate more functions $f$ that better correlates with the random noise on average.

**Theorem 2.** *Based on the TTN-VQC setup in Theorem 1, the estimation error is upper bounded by the empirical Rademacher complexity $2\hat{\mathcal{R}}_S(\mathbb{F}_{TV})$, which is*

$$2\hat{\mathcal{R}}_S(\mathbb{F}_{TV}) \leq 2\hat{\mathcal{R}}_S(\mathbb{F}_{TTN}) + 2\hat{\mathcal{R}}_S(\mathbb{F}_{VQC}) \leq \frac{2P}{\sqrt{N}}\sqrt{\sum_{k=1}^{K}\Lambda_k^2} + \frac{2P\Lambda'}{\sqrt{N}}$$

$$s.t., \|\boldsymbol{x}_n\|_2 \leq P, \;\; \forall n \in [N], \tag{13}$$

$$\|\boldsymbol{W}(\mathcal{T}_{\boldsymbol{\theta}_{vqc}})\|_F \leq \Lambda', \;\; \|\mathcal{W}^{[k]}(\mathcal{T}_{\boldsymbol{\theta}_{ttn}})\|_F \leq \Lambda_k, \; k \in [K],$$

*where $\mathbb{F}_{TTN}$ and $\mathbb{F}_{VQC}$ separately denote the family of TTN and VQC, $P$, $\Lambda'$ and $\Lambda_k$ are constants, $\boldsymbol{W}(\mathcal{T}_{\boldsymbol{\theta}_{vqc}})$ refers to a matrix associated with the operator $\mathcal{T}_{\boldsymbol{\theta}_{vqc}}$, and $\mathcal{W}^{[k]}(\mathcal{T}_{\boldsymbol{\theta}_{ttn}})$ corresponds to a 4-order tensor of TTN, $\|\boldsymbol{W}\|_F$ and $\|\mathcal{W}^{[k]}\|_F$ represent the Frobenius norm of a matrix and a tensor, respectively.*

The upper bound on the estimation error in Eq. (13) shows when an input $\mathbf{x}$ and an initialized TTN-VQC model are given, a sufficiently large amount of training data $N$ is needed to lower the related upper bound. On the other hand, the noise perturbation associated with the noisy power $P_{noise}$ imposed upon the input corresponds to a larger total power $P = P_{in} + P_{noise}$, which corresponds to a larger upper bound on the estimation error and accordingly weakens the generalization power.

### 2.2.4 Upper Bounds on Optimization Error

A QNN system always suffers from the problem of Barren Plateaus [31], which results from optimizing a non-convex objective function and the gradients may vanish almost everywhere in the training stage. To alleviate the problem of Barren Plateaus, we introduce a new initialization strategy based on the Polyak-Lojasiewicz (PL) condition [47, 48, 49]. More specifically, given the set of model parameters $\boldsymbol{\theta} = \{\boldsymbol{\theta}_{ttn}, \boldsymbol{\theta}_{vqc}\}$ for TTN-VQC, if an empirical loss function $\mathcal{L}_S$ satisfies $\mu$-PL, the $L_2$-norm of the first-order gradient $\nabla\mathcal{L}_S$ concerning $\boldsymbol{\theta}$ should satisfy the following inequality as:

$$\frac{1}{2}\|\nabla\mathcal{L}_S(\boldsymbol{\theta})\|_2^2 \geq \mu\mathcal{L}_S(\boldsymbol{\theta}). \tag{14}$$

**Theorem 3.** *If a 1-Lipschitz loss function $\mathcal{L}$ over the set of TTN-VQC parameters $\boldsymbol{\theta}$ satisfies the PL condition, the gradient descent algorithm with a learning rate of 1 can lead to an exponential convergence rate. More specifically, at epoch $T$, we have*

$$\mathcal{L}_S(\boldsymbol{\theta}_T) \leq \exp\left(-\mu T\right)\mathcal{L}_S(\boldsymbol{\theta}_0), \tag{15}$$

*where $\boldsymbol{\theta}_0$ and $\boldsymbol{\theta}_T$ separately denote the parameters at the initial stage and at the epoch T. Furthermore, given a radius $r = 2\sqrt{2\mathcal{L}_S(\boldsymbol{\theta}_0)}/\mu$ for a closed ball $B(\boldsymbol{\theta}_0, r)$, there exists a global minimum hypothesis $\boldsymbol{\theta}^* \in B(\boldsymbol{\theta}_0, r)$ such that the optimization error becomes sufficiently small.*

Furthermore, we show a necessary condition in Proposition 2 for a TTN-VQC operator $f \in \mathbb{F}_{TV}$ to satisfy the $\mu$-PL setup of $\mathcal{L}_S(\boldsymbol{\theta})$, which is related to the tangent kernel of the operator $f$.

**Proposition 2.** *For a TTN-VQC operator $f \in \mathbb{F}_{TV}$, we define the tangent kernel $\mathcal{K}_f$ as $\nabla f(\boldsymbol{\theta})\nabla f(\boldsymbol{\theta})^T$. If a 1-Lipschitz loss function $\mathcal{L}_S(\boldsymbol{\theta})$ satisfies the $\mu$-PL condition, $\lambda_{\min}(\mathcal{K}_f)$ represents the smallest eigenvalue of $\mathcal{K}_f$ and meets the condition as:*

$$\lambda_{\min}(\mathcal{K}_f) \geq \mu. \tag{16}$$

Theorem 3 suggests that the $\mu$-PL condition for the TTN-VQC ensures an exponential convergence rate and the training loss can reach as low as 0. Proposition 2 can check if the $\mu$-PL condition can be met by calculating its tangent kernel. Our theorems suggest that the TTN-VQC model meeting the PL condition can better deal with the problem of Barren Plateaus, but we cannot guarantee that the model with a low optimization bias has to meet the PL condition. In other words, the PL condition is one of the potential approaches to ensure the VQC handles the optimization issue.

### 2.2.5 Putting It All Together

Based on the derived upper bound, under the setup of $\mu$-PL condition, the upper bounds on the error components can be combined into an aggregated upper bound as:

$$
\begin{aligned}
\mathcal{L}_{\mathcal{D}}(\bar{f}_S) &\leq \mathcal{L}_{\mathcal{D}}(f_{\mathcal{D}}^*) + 2\hat{\mathcal{R}}_{\mathcal{S}}(\mathbb{F}_{TV}) + \nu \\
&\leq \frac{\Theta(1)}{\sqrt{U}} + \mathcal{O}\left(\frac{1}{\sqrt{M}}\right) + \frac{2P}{\sqrt{N}}\sqrt{\sum_{k=1}^{K}\Lambda_k^2} + \frac{2P\Lambda'}{\sqrt{N}} \\
& s.t., \quad \|\mathbf{x}_n\|_2 \leq P, \ n \in [N], \\
& \|\mathbf{W}(\mathcal{T}_{\boldsymbol{\theta}_{vqc}})\|_F \leq \Lambda', \ \|\mathcal{W}^{[k]}(\mathcal{T}_{\boldsymbol{\theta}_{ttn}})\|_F \leq \Lambda_k, k \in [K].
\end{aligned}
\tag{17}
$$

The aggregated upper bound in Eq. (17) shows that the training error $\epsilon$ can be reduced to closely 0 with the setup of $\mu$-PL condition, and the expected loss is mainly determined by the upper bounds on the approximation and estimation errors.

### 2.3 Empirical Results

To separately corroborate our theoretical analysis of the TTN-VQC, our experiments are composed of two groups: (1) to evaluate the representation power, the training and test datasets are set in the same clean environment; (2) to assess the generalization power of TTN-VQC, the test data are separately mixed by additive Gaussian and Laplacian noises, where the SNR levels are set as 8dB and 12dB, respectively. Our baseline system is a linear PCA-VQC model where the technique of principal component analysis (PCA) [50] is employed. PCA is a standard method to reduce data dimensionality by using a linear transformation in an unsupervised manner. Our experiments compare the performance of the TTN-VQC and PCA-VQC models, and particularly aim at verifying the following points:

1. The TTN-VQC can lead to better performance than PCA-VQC in both matched and unmatched environmental settings.

2. Increasing the number of qubits can improve the representation power of TTN-VQC.

3. Exponential convergence rates demonstrate our configurations of the TTN-VQC model satisfy the $\mu$-PL condition.

We evaluate the performance of TTN-VQC on the standard MNIST dataset [51]. The MNIST dataset aims at the task of handwritten 10 digit classification, where there are $60,000$ examples for training and $10,000$ data for testing. In our experiments, we randomly sample $10,000$ in training data and $2,000$ in test data. Both training and test data are corrupted with noisy signals at different SNR levels, and the generated noisy data are taken as the input to the quantum-based models. The target of the models is set as the clean data during the training stage, where the model-enhanced data are expected to be as close as the target one. We measure the model performance in the test stage by calculating the $L_1$-norm loss between enhanced data and target one.

As the experimental baseline, a hybrid PCA-VQC model is conducted, where PCA serves as a simple feature extractor followed by the VQC as the classifier. The PCA-VQC represents a linear VQC model which is in contrast to a nonlinear one based on the TTN-VQC model. We include 4 PQC blocks in the VQC employed in the experiments. As for the experiments of TTN-VQC, the image data are reshaped into a 3-order $7 \times 16 \times 7$ tensors. Given a set of ranks $\mathbf{R} = \{1, 3, 3, 1\}$, we can set 3 trainable tensors as: $\mathcal{W}_1 \in \mathbb{R}^{1 \times 7 \times U_1 \times 3}$, $\mathcal{W}_2 \in \mathbb{R}^{3 \times 16 \times U_2 \times 3}$, and $\mathcal{W}_3 \in \mathbb{R}^{3 \times 7 \times U_3 \times 1}$, where $U = \prod_{k=1}^{3} U_k$ is associated with the number of qubits. In particular, we separately assess the models with 8 qubits and 12 qubits, and the parameters $(U_1, U_2, U_3)$ are set as $(2, 2, 2)$ for the 8 qubits and $(2, 3, 2)$ for the 12 qubits. The stochastic gradient descent (SGD) [52] with an Adam optimizer [53] is utilized in the training process, where a mini-batch of 50 and a learning rate of 1 are configured. The 1-Lipschitz continuous function based on MAE is taken to meet the PL condition.

### 2.3.1 Experiments for Representation Power of TTN-VQC

To corroborate the Theorem 1 for the representation power of TTN-VQC, both training and test data are mixed with the Gaussian noise of the 15dB SNR level, and we compare the performance of TTN-VQC with PCA-VQC on the generated noisy settings. Figure 5 demonstrate the related empirical results, where TTN-VQC_8Qubit and TTN-VQC_12Qubit separately represent the TTN-VQC models with 8 and 12 qubits and PCA-VQC_8Qubit and PCA-VQC_12Qubit denote that the PCA-VQC models with 8 and 12 qubits, respectively. Our experiments show that the TTN-VQC can significantly outperform the PCA-VQC counterparts in terms of lower training and test loss values. Moreover, our results also suggest that more qubits can improve the empirical performance of both TTN-VQC and PCA-VQC models. Table 2 presents the final results on the test dataset. The TTN-VQC_12Qubit model owns more parameters than the TTN-VQC_8Qubit model (0.636Mb vs. 0.452Mb), but the former one attains better empirical performance in terms of lower MAE scores (0.0156 vs. 0.0597) on the test dataset.

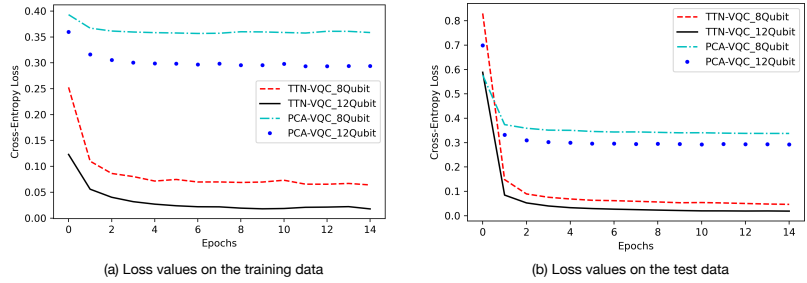

(a) Loss values on the training data          (b) Loss values on the test data

Figure 5: *Empirical results of the vector-to-vector regression on the MNIST dataset to evaluate the representation power of TTN-VQC. TTN-VQC_8Qubit and TTN-VQC_12Qubit represent the TTN-VQC models with 8 and 12 qubits, respectively; PCA-VQC_8Qubit and PCA-VQC_12Qubit separately denote the PCA-VQC models with 8 and 12 qubits.*

| Models | Qubits | Params (Mb) | MAE |
|---|---|---|---|
| TTN-VQC_8Qubit | 8 | 0.452 | 0.0597 |
| TTN-VQC_12Qubit | 12 | 0.636 | 0.0156 |
| PCA-VQC_8Qubit | 8 | 0.080 | 0.3847 |
| PCA-VQC_12Qubit | 12 | 0.120 | 0.2939 |

Table 2: Empirical results of TTN-VQC and PCA-VQC models on the test dataset.

### 2.3.2 Experiments for Generalization Power of TTN-VQC

To assess the generalization power of TTN-VQC, the test data are separately mixed with additive Gaussian and Laplacian noises with 8dB and 12dB SNR levels. Based on the well-trained TTN-VQC and PCA-VQC models with 8 qubits, we further assess their performance on the test data with Gaussian and Laplacian noisy conditions related to the evaluation of their generalization power. Based on the upper bound of the generalization power in Theorem 2, given the input dataset, a more noisy setting corresponds to a larger $P_{noise}$, which results in a larger total power $P = P_{in} + P_{noise}$. Thus, we corroborate our theorem in the experiment by evaluating the empirical performance under different noisy conditions. In the meanwhile, to highlight the advantage of non-linearity for TTN-VQC, we also compare the experimental results of both TTN-VQC and PCA-VQC.

For one thing, Figure 6 suggests that the TTN-VQC models significantly outperform the PCA-VQC counterparts in the two noisy settings, and Table 3 shows the MAE scores of TTN-VQC and PCA-VQC models, where the TTN-VQC models achieve much better performance than the PCA-VQC ones in terms of lower MAE scores under all kinds of noisy environments. For another, we observe that the experimental performance of the TTN-VQC models under more adverse Gaussian and Laplacian noisy settings is degraded because of higher MAE scores, which corresponds to our theoretical analysis.

Moreover, our derived upper bound on the estimation error is also associated with the amount of training data. To test the effect of training data for the generalization capability, the number of training data is gradually incremented from a subset of data to a whole set. In Table 4, we observe that a larger amount of training data leads to lower MAE scores which correspond to better generalization power.

| Models | Noise Type | Params (Mb) | MAE |
|--------|-----------|-------------|-----|
| TTN-VQC_8Qubit | Gaussian (8dB) | 0.452 | 0.1703 |
| TTN-VQC_8Qubit | Gaussian (12dB) | 0.452 | 0.1078 |
| PCA-VQC_8Qubit | Gaussian (8dB) | 0.080 | 0.5151 |
| PCA-VQC_8Qubit | Gaussian (12dB) | 0.080 | 0.4546 |
| TTN-VQC_8Qubit | Laplacian (8dB) | 0.452 | 0.1684 |
| TTN-VQC_8Qubit | Laplacian (12dB) | 0.452 | 0.1327 |
| PCA-VQC_8Qubit | Laplacian (8dB) | 0.080 | 0.4651 |
| PCA-VQC_8Qubit | Laplacian (12dB) | 0.080 | 0.4396 |

Table 3: Empirical results of TTN-VQC and PCA-VQC models on the test dataset with either Gaussian or Laplacian noise with 8dB or 12dB SNR levels.

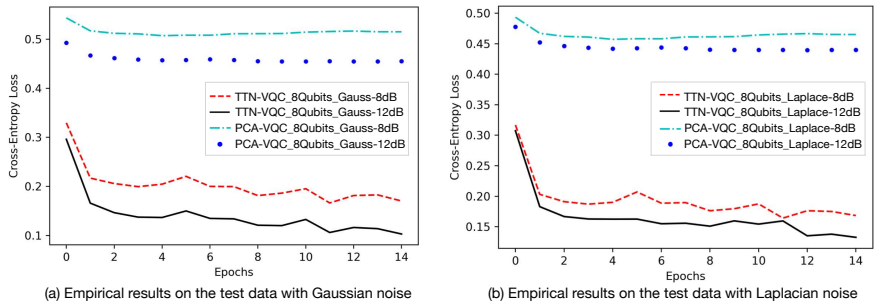

Figure 6: *Empirical results of the vector-to-vector regression on the MNIST dataset to evaluate the generalization power of TTN-VQC and PCA-VQC with 8 qubits. There are two noisy settings on the test dataset to evaluate the performance of the TTN-VQC and PCA-VQC models: (a) Gauss-8dB and Gauss-12dB separately denote the Gaussian noisy conditions of 8dB and 12dB SNR levels; (b) Laplace-8dB and Laplace-12dB refer to the Laplacian noisy settings of 8dB and 12dB SNR levels, respectively.*

## 3   DISCUSSION

This work focuses on the theoretical error performance analysis for VQC-based functional regression, particularly when the TTN is employed for dimensionality reduction. Our theoretical results provide upper bounds on the representation and generalization powers of TTN-VQC. Our theoretical results suggest that the approximation error is inversely proportional to the square root of qubits, which means that the increase of qubits can lead to better representation power of TTN-VQC. The estimation error of TTN-VQC is related to its generalization power, which is upper-bounded based on the empirical Rademacher complexity. The optimization error can be lowered to a small score by leveraging the PL condition to realize an exponential convergence based on the SGD algorithm. To our best knowledge, no prior works, such as a complete error characterization, have been delivered.

Our experiments of vector-to-vector regression on the MNIST dataset are designed to corroborate the theoretical results. We first compare the representation power of the TTN-VQC models with the PCA-VQC counterparts. We observe that more qubits and the non-linear property for TTN-VQC can improve the empirical performance that matches our theoretical analysis. Further, we assess the generalization power of TTN-VQC by taking different noisy inputs into account, and we demonstrate that more mismatched and noisy inputs can worsen the generalization power. Besides, the non-linear TTN-VQC models outperform the linear PCA-VQC models in terms of representation and generalization powers. That implies that the non-linearity of TTN-VQC can greatly contribute to the improvement of VQC performance.

| Models | Noise Type | Amount of Training Data | MAE |
|--------|-----------|------------------------|-----|
| TTN-VQC_8Qubit | Gaussian (12dB) | 20,000 | 0.2941 |
| TTN-VQC_8Qubit | Gaussian (12dB) | 40,000 | 0.1853 |
| PCA-VQC_8Qubit | Gaussian (12dB) | 60,000 | 0.1078 |

Table 4: Empirical results of TTN-VQC on datasets of different sizes. Three groups of data sizes are attempted and a large amount of training data achieves a lower MAE score.

We also note that the TTN-VQC models attain exponential convergence rates. The optimization error is eventually reduced to $0$ in the training process, which corresponds to the PL condition in our theoretical analysis. Moreover, the empirical results on the test dataset consistently exhibit a decreasing trend. The empirical results imply that the model setup for TTN-VQC meets the PL condition and thus it can handle the problem of Barren Plateaus. Our future work will discuss how to initialize the VQC model based on the PL condition to minimize the optimization bias.

Furthermore, our theoretical results are built upon the Lipschitz loss function utilized for the regression problem, and the theoretical contributions can be certainly generalized to the classification tasks where the loss functions like hinge loss and cross-entropy are data-dependent Lipschitz continuity and the Lipschitz constant does not keep the same value on different datasets.

# 4 METHOD

This section aims at providing detailed proof of our theoretical results. We first present the upper bound on the representation power, and then we derive another upper bound on the generalization power. The analysis of optimization performance is also conducted based on the PL condition.

## 4.1 Proof for Theorem 1

The derivation of Theorem 1 is mainly based on the classical universal approximation theorem [44, 45, 46] and a parallel structure of TTN. We first assume $g_m(\mathbf{x}; \boldsymbol{\theta}_{vqc}, \boldsymbol{\theta}_{ttn})$ as the $m$-th measurement for the TTN-VQC operator $g(\mathbf{x}; \boldsymbol{\theta}_{vqc}, \boldsymbol{\theta}_{ttn})$, and $\sum_{m=1}^{M} g_m(\mathbf{x}; \boldsymbol{\theta}_{vqc}, \boldsymbol{\theta}_{ttn})$ is defined as:

$$
\begin{aligned}
\sum_{m=1}^{M} g_m(\mathbf{x}; \boldsymbol{\theta}_{vqc}, \boldsymbol{\theta}_{ttn}) &= \sum_{m=1}^{M} \mathcal{M}_m \circ \mathcal{T}_{\boldsymbol{\theta}_{vqc}} \circ \mathcal{T}_{\mathbf{y}} \circ \mathcal{T}_{\boldsymbol{\theta}_{ttn}}(\mathbf{x}) \\
&= \mathcal{M}' \circ \mathcal{T}_{\boldsymbol{\theta}_{vqc}} \circ \mathcal{T}_{\mathbf{y}} \circ \mathcal{T}_{\boldsymbol{\theta}_{ttn}}(\mathbf{x}) \\
&= \mathcal{M}' \circ \mathcal{H} \circ \mathcal{T}_{\boldsymbol{\theta}_{ttn}}(\mathbf{x}),
\end{aligned}
$$

where the operator $\mathcal{H} = \mathcal{T}_{\boldsymbol{\theta}_{vqc}} \circ \mathcal{T}_{\mathbf{y}}$ refers to a unitary matrix, and $\mathcal{M}_m$ denotes the $m$-th measurement and $\mathcal{M}' = \sum_{m=1}^{M} \mathcal{M}_m$. Moreover, $\mathcal{H}^{-1}$ is a reversely linear unitary operator of $\mathcal{H}$, and $g_m$ refers to the function after the quantum measurement. Next, we can further derive that

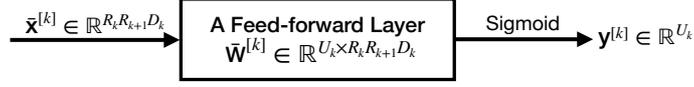

Figure 7: *The $k^{th}$ channel of TTN is equivalent to a feed-forward layer of neural network with the sigmoid function. The input $\bar{\mathbf{x}}^{[k]}$ is derived from the reshape of $\mathcal{X}^{[k]}$ and goes through the a feed-forward neural network with the weight matrix $\bar{\mathbf{W}}^{[k]}$ and sigmoid function. The output $\mathbf{y}^{[k]}$ corresponds to the array for the $k^{th}$ order of $\hat{\mathcal{Y}}$.*

$$
\left\| \hat{f}(\mathbf{x}) - \mathcal{T}_{lr}(\mathbb{E}[g(\mathbf{x}; \boldsymbol{\theta}_{vqc}, \boldsymbol{\theta}_{ttn})]) \right\|_1
$$

$$
\leq \left\| \hat{f}(\mathbf{x}) - \mathcal{T}_{lr}\left( \frac{1}{M} \sum_{m=1}^{M} g_m(\mathbf{x}; \boldsymbol{\theta}_{vqc}, \boldsymbol{\theta}_{ttn}) \right) \right\|_1 \qquad \text{(Triangle Ineq.)}
$$

$$
+ \left\| \mathcal{T}_{lr}\left( \frac{1}{M} \sum_{m=1}^{M} g_m(\mathbf{x}; \boldsymbol{\theta}_{vqc}, \boldsymbol{\theta}_{ttn}) \right) - \mathcal{T}_{lr}\left( \mathbb{E}[g(\mathbf{x}; \boldsymbol{\theta}_{vqc}, \boldsymbol{\theta}_{ttn})] \right) \right\|_1
$$

$$
= \left\| \mathcal{T}_{lr}\left( \mathcal{T}_{lr}^{-1}(\hat{f}(\mathbf{x})) - \frac{1}{M} \sum_{m=1}^{M} g_m(\mathbf{x}; \boldsymbol{\theta}_{vqc}, \boldsymbol{\theta}_{ttn}) \right) \right\|_1
$$

$$
+ \left\| \mathcal{T}_{lr}\left( \frac{1}{M} \sum_{m=1}^{M} g_m(\mathbf{x}; \boldsymbol{\theta}_{vqc}, \boldsymbol{\theta}_{ttn}) - \mathbb{E}[g(\mathbf{x}; \boldsymbol{\theta}_{vqc}, \boldsymbol{\theta}_{ttn})] \right) \right\|_1
$$

$$
\leq \left\| \mathcal{T}_{lr}\left( \mathcal{M}' \circ \mathcal{H} \circ \mathcal{H}^{-1} \circ \mathcal{T}_{lr}^{-1}(\hat{f}(\mathbf{x})) - \mathcal{M}' \circ \mathcal{H} \circ \mathcal{T}_{\boldsymbol{\theta}_{ttn}}(\mathbf{x}) \right) \right\|_1
$$

$$
+ \mathcal{O}\left( \frac{1}{\sqrt{M}} \right) \cdot \|\mathcal{T}_{lr}(1)\|_1 \qquad \text{(Central Limit Theorem)}
$$

$$
\leq \prod_{k=1}^{K} \frac{1}{\sqrt{U_k}} \cdot \|\mathcal{T}_{lr} \circ \mathcal{M}' \circ \mathcal{H}(1)\|_1 + \mathcal{O}\left( \frac{1}{\sqrt{M}} \right) \cdot \|\mathcal{T}_{lr}(1)\|_1 \quad \text{(Universal Approx.)}
$$

$$
= \frac{\Theta(1)}{\sqrt{U}} + \mathcal{O}\left( \frac{1}{\sqrt{M}} \right) \qquad\qquad (\prod_{k=1}^{K} U_k = U).
$$

## 4.2 Proof for Theorem 2

Based on Eq. (9) and Figure 4, the $k^{th}$ channel is equivalent to a feed-forward layer of neural network with the sigmoid function. More specifically, the input $\mathcal{X}^{[k]} \in \mathbb{R}^{R_k \times D_k \times R_{k+1}}$ is reshaped into a high-dimensional vector $\bar{\mathbf{x}}^{[k]} \in \mathbb{R}^{R_k R_{k+1} D_k}$, which further goes through the feed-forward layer with the weight matrix $\bar{\mathbf{W}}^{[k]} \in \mathbb{R}^{U_k \times R_k R_{k+1} D_k}$. After the operation of sigmoid function, we have an output vector $\mathbf{y}^{[k]} \in \mathbb{R}^{U_k}$.

As for the upper bound for the TTN-VQC model on the estimation error, we separately upper bound each term of the TTN and VQC families by leveraging the empirical Rademacher complexity. Moreover, we define $\hat{\mathcal{R}}_S(\mathbb{F}_{TTN}^{[k]})$ as the functional family for the $k^{th}$ channel associated with Figure 7.

Thus, based on the Rademacher identities, we attain that $\hat{\mathcal{R}}_S(\mathbb{F}_{TTN}) \leq \sum_{k=1}^{K} \hat{\mathcal{R}}_S(\mathbb{F}_{TTN}^{[k]})$.

$$
\begin{aligned}
\hat{\mathcal{R}}_S(\mathbb{F}_{TTN}^{[k]}) &= \frac{1}{N}\mathbb{E}_{\boldsymbol{\epsilon}}\left[\sup_{||\bar{\mathbf{w}}_u^{[k]}||_2 \leq \Lambda} \sum_{n=1}^{N} \epsilon_n \sum_{u=1}^{U} \sigma(\bar{\mathbf{w}}_u^{[k]} \cdot \bar{\mathbf{x}}_n^{[k]})\right] \\
&= \frac{1}{N}\mathbb{E}_{\boldsymbol{\epsilon}}\left[\sup_{||\bar{\mathbf{w}}_u^{[k]}||_2 \leq \Lambda} \sum_{u=1}^{U} \sum_{n=1}^{N} \epsilon_n \sigma(\bar{\mathbf{w}}_u^{[k]} \cdot \bar{\mathbf{x}}_n^{[k]})\right] \\
&= \frac{1}{N}\mathbb{E}_{\boldsymbol{\epsilon}}\left[\sup_{||\bar{\mathbf{w}}_u^{[k]}||_2 \leq \Lambda, u \in [1,U]} \left|\sum_{n=1}^{N} \epsilon_n \sigma(\bar{\mathbf{w}}_u^{[k]} \cdot \bar{\mathbf{x}}_n^{[k]})\right|\right] \\
&= \frac{1}{N}\mathbb{E}_{\boldsymbol{\epsilon}}\left[\sup_{||\bar{\mathbf{w}}_u^{[k]}||_2 \leq \Lambda, u \in [1,U]} \sup_{s \in \{-1,+1\}} s \sum_{n=1}^{N} \epsilon_n \sigma(\bar{\mathbf{w}}_u^{[k]} \cdot \bar{\mathbf{x}}_n^{[k]})\right].
\end{aligned}
\tag{18}
$$

Furthermore, we upper bound $\hat{\mathcal{R}}_S(\mathbb{F}_{TTN}^{[k]})$ by utilizing Talagrand inequality [54] and we obtain

$$
\begin{aligned}
\hat{\mathcal{R}}_S(\mathbb{F}_{TTN}^{[k]}) &\leq \frac{1}{N}\mathbb{E}_{\boldsymbol{\epsilon}}\left[\sup_{||\bar{\mathbf{w}}_u^{[k]}||_2 \leq \Lambda, u \in [1,U]} \sup_{s \in \{-1,+1\}} s \sum_{n=1}^{N} \epsilon_n \bar{\mathbf{w}}_u^{[k]} \cdot \bar{\mathbf{x}}_n^{[k]}\right] \\
&= \frac{1}{N}\mathbb{E}_{\boldsymbol{\epsilon}}\left[\sup_{||\bar{\mathbf{w}}_u^{[k]}||_2 \leq \Lambda, u \in [1,U]} \left|\bar{\mathbf{w}}_u^{[k]} \cdot \sum_{n=1}^{N} \epsilon_n \bar{\mathbf{x}}_n^{[k]}\right|\right] \\
&= \frac{\Lambda_k}{N}\mathbb{E}_{\boldsymbol{\epsilon}}\left[\left|\left|\sum_{n=1}^{N} \epsilon_n \bar{\mathbf{x}}_n^{[k]}\right|\right|_2\right] \\
&\leq \frac{\Lambda_k}{N}\sqrt{\mathbb{E}_{\boldsymbol{\epsilon}}\left[\left|\left|\sum_{n=1}^{N} \epsilon_n \bar{\mathbf{x}}_n^{[k]}\right|\right|_2^2\right]} \qquad \text{(Jensen's inequality)} \\
&= \frac{\Lambda_k}{N}\sqrt{\sum_{i,j=1}^{N} \mathbb{E}_{\boldsymbol{\epsilon}}[\epsilon_i \epsilon_j](\bar{\mathbf{x}}_i^{[k]} \cdot \bar{\mathbf{x}}_j^{[k]})} \\
&= \frac{\Lambda_k}{N}\sqrt{\sum_{i,j=1}^{N} 1_{i=j}(\bar{\mathbf{x}}_i^{[k]} \cdot \bar{\mathbf{x}}_j^{[k]})} \\
&= \frac{\Lambda_k}{N}\sqrt{\sum_{n=1}^{N} \left|\left|\bar{\mathbf{x}}_n^{[k]}\right|\right|_2^2} \qquad (||\bar{\mathbf{x}}_n^{[k]}||_2^2 \leq P_k^2) \\
&\leq \frac{\Lambda_k P_k}{\sqrt{N}},
\end{aligned}
\tag{19}
$$

where we assume $||\bar{\mathbf{x}}_n^{[k]}||_2 \leq P_k$ and accordingly $\sqrt{\sum_{n=1}^{N} ||\mathbf{x}_n^{[k]}||_2^2} \leq \sqrt{N}P_k$.

Finally, we utilize the Cauchy-Schwarz inequality and obtain the result that

$$
\hat{\mathcal{R}}_S(\mathbb{F}_{TTN}) \leq \sum_{k=1}^{K} \hat{\mathcal{R}}_S(\mathbb{F}_{TTN}^{[k]}) = \sum_{k=1}^{K} \frac{\Lambda_k P_k}{\sqrt{N}} \leq \frac{\sqrt{\sum_{k=1}^{K} \Lambda_k^2}\sqrt{\sum_{k=1}^{K} P_k^2}}{\sqrt{N}},
\tag{20}
$$

where $P = \sqrt{\sum_{k=1}^{K} P_k^2}$ and $||\mathbf{x}_n||_2 \leq \sum_{k=1}^{K} \bar{\mathbf{x}}_n^{[k]} = \sum_{k=1}^{K} P_k \leq \sqrt{\sum_{k=1}^{K} P_k^2} = P$. Hence, we attain the inequality as follows:

$$
\hat{\mathcal{R}}_S(\mathbb{F}_{TTN}) \leq \frac{P\sqrt{\sum_{k=1}^{K} \Lambda_k^2}}{\sqrt{N}},
\tag{21}
$$

$$
s.t., ||\mathbf{x}_n||_2 \leq P,\, n \in [N],\, ||\mathcal{W}^{[k]}(\mathcal{T}_{\boldsymbol{\theta}_{ttn}})||_F \leq \Lambda_k,\, k \in [K].
$$

Similarly, we can also obtain the result that $\hat{\mathcal{R}}_S(\mathbb{F}_{VQC}) \leq \frac{P\Lambda'}{\sqrt{N}}$ with the constraint that $||\mathbf{W}(\mathcal{T}_{\boldsymbol{\theta}_{vqc}})||_F \leq \Lambda'$. Then, we complete the proof for Theorem 2.

### 4.3 Proof for Theorem 3

Assume the gradient descent algorithm runs around the closed ball $B(\boldsymbol{\theta}_0, R)$ with $R = \frac{2}{\mu}$ and the loss function $\mathcal{L}_S(\boldsymbol{\theta})$ has the following properties as: (1) The loss $\mathcal{L}_S(\boldsymbol{\theta})$ is $\mu$-PL; (2) The loss $\mathcal{L}_S(\boldsymbol{\theta})$ is 1-Lipschitz; (3) The norm of Hessian $H$ is bounded by 1.

Then, we need to prove the following properties: (a) There exists a global minimum $\boldsymbol{\theta}^* \in B(\boldsymbol{\theta}_0, R)$; (b) The algorithm of gradient descent converges with an exponential convergence rate: $\mathcal{L}_S(\boldsymbol{\theta}_{t+1}) \leq (1 - \eta\mu)^{t+1}\mathcal{L}_S(\boldsymbol{\theta}_0)$. By applying the Taylor expansion, we obtain

$$
\begin{aligned}
&\mathcal{L}_S(\boldsymbol{\theta}_{t+1}) \\
&= \mathcal{L}_S(\boldsymbol{\theta}_t) + (\boldsymbol{\theta}_{t+1} - \boldsymbol{\theta}_t)^T \nabla f(\boldsymbol{\theta}_t) + \frac{1}{2}(\boldsymbol{\theta}_{t+1} - \boldsymbol{\theta}_t)^T H(\boldsymbol{\theta}')(\boldsymbol{\theta}_{t+1} - \boldsymbol{\theta}_t) \\
&= \mathcal{L}_S(\boldsymbol{\theta}_t) + (-\eta)\nabla\mathcal{L}_S(\boldsymbol{\theta}_t)^T \nabla\mathcal{L}_S(\boldsymbol{\theta}_t) + \frac{1}{2}(-\eta)\nabla\mathcal{L}_S(\boldsymbol{\theta}_t)^T H(\boldsymbol{\theta}')(-\eta)\nabla\mathcal{L}_S(\boldsymbol{\theta}_t) \\
&= \mathcal{L}_S(\boldsymbol{\theta}_t) - \eta\|\nabla\mathcal{L}_S(\boldsymbol{\theta}_t)\|_2^2 + \frac{\eta^2}{2}\nabla\mathcal{L}_S(\boldsymbol{\theta}_t)^T H(\boldsymbol{\theta}')\nabla\mathcal{L}_S(\boldsymbol{\theta}_t) \\
&\leq \mathcal{L}_S(\boldsymbol{\theta}_t) - \eta(1 - \frac{\eta}{2})\|\nabla\mathcal{L}_S(\boldsymbol{\theta}_t)\|_2^2 && \text{(by Assumption 3)} \\
&\leq \mathcal{L}_S(\boldsymbol{\theta}_t) - \eta(2 - \eta)\mu\mathcal{L}_S(\boldsymbol{\theta}_t) && \text{(by } \mu\text{-PL Assumption)} \\
&= \left(1 - 2\eta\mu + \eta^2\mu\right)\mathcal{L}_S(\boldsymbol{\theta}_t) \\
&\leq \left(1 - 2\eta\mu + \eta^2\mu\right)^{t+1}\mathcal{L}_S(\boldsymbol{\theta}_0).
\end{aligned}
$$

Next, we show that $\boldsymbol{\theta}_t$ does not leave the ball $B$. Based on the assumption 4, we have $\mathcal{L}(\boldsymbol{\theta}_t) - \mathcal{L}(\boldsymbol{\theta}_{t+1}) \geq \frac{1}{2}\|\nabla\mathcal{L}(\boldsymbol{\theta}_t)\|_2^2$, which leads to $\|\nabla\mathcal{L}(\boldsymbol{\theta}_t)\|_2^2 \leq \sqrt{2\beta(\mathcal{L}(\boldsymbol{\theta}_t) - \mathcal{L}(\boldsymbol{\theta}_{t+1}))}$. Then, we further derive that

$$
\begin{aligned}
||\boldsymbol{\theta}_{t+1} - \boldsymbol{\theta}_0||_2^2 &= \eta \left\|\sum_{\tau=0}^{t} \nabla\mathcal{L}(\boldsymbol{\theta}_\tau)\right\|_2^2 \\
&\leq \eta \sum_{\tau=0}^{t} \|\nabla\mathcal{L}(\boldsymbol{\theta}_\tau)\|_2^2 \\
&\leq \eta \sum_{\tau=0}^{t} \sqrt{2\left(\mathcal{L}(\boldsymbol{\theta}_\tau) - \mathcal{L}(\boldsymbol{\theta}_{\tau+1})\right)} && \text{(by Continuity)} \\
&\leq \eta \sum_{\tau=0}^{t} \sqrt{2\mathcal{L}(\boldsymbol{\theta}_\tau)} \\
&\leq \eta\sqrt{2} \sum_{\tau=0}^{t} \sqrt{(1 - 2\eta\mu + \eta^2\mu)^\tau \mathcal{L}(\boldsymbol{\theta}_0)} && \text{(by Geometric Convergence)} \\
&= \eta\sqrt{2\mathcal{L}(\boldsymbol{\theta}_0)} \sum_{\tau=0}^{t} (1 - 2\eta\mu + \eta^2\mu)^{\tau/2} \\
&= \frac{\eta\sqrt{2\mathcal{L}(\boldsymbol{\theta}_0)}}{1 - \sqrt{1 - 2\eta\mu + \eta^2\mu}} \\
&= \frac{\sqrt{2\mathcal{L}(\boldsymbol{\theta}_0)}(1 + \sqrt{1 - 2\eta\mu + \eta^2\mu})}{\mu(2 - \eta)} \\
&\leq \frac{2\sqrt{2\mathcal{L}(\boldsymbol{\theta}_0)}}{\mu} && \text{(By Setting } \eta = 1\text{).}
\end{aligned}
$$

The inequality $\|\boldsymbol{\theta}_{t+1} - \boldsymbol{\theta}_0\|_2^2 \leq \frac{2\sqrt{2\mathcal{L}(\boldsymbol{\theta}_0)}}{\mu}$ represents the gradient descent algorithm ensures the updated point in a ball with a radius of $\frac{2\sqrt{2\mathcal{L}(\boldsymbol{\theta}_0)}}{\mu}$, and a larger $\mu$ leads to larger updates and faster convergence rate over a smaller ball.

## COMPETING INTERESTS

The authors declare no Competing Financial or Non-Financial Interests.

## AUTHOR CONTRIBUTIONS